\documentclass[apj]{emulateapj}
\usepackage{natbib}
\usepackage{amstext}
\usepackage{graphicx}

\newcommand{\pder}[2]{\frac{\partial#1}{\partial#2}}

\begin{document}

\title{Correcting for interplanetary scattering in velocity dispersion
analysis of solar energetic particles}
\shorttitle{Velocity Dispersion and Solar Energetic Particles}

\author{T. Laitinen}
\affil{Jeremiah Horrocks Institute, University of Central Lancashire, UK}
\email{timo.lm.laitinen@gmail.com}
\author{K. Huttunen-Heikinmaa and E. Valtonen}
\affil{Department of Physics, University of Turku, Finland}
\author{S. Dalla}
\affil{Jeremiah Horrocks Institute, University of Central Lancashire, UK}

\shortauthors{Laitinen et al.}

\begin{abstract}
  To understand the origin of Solar Energetic Particles (SEPs),
  we must study their injection time relative to other solar eruption
  manifestations. Traditionally the injection time is determined using
  the Velocity Dispersion Analysis (VDA) where a linear fit of the
  observed event onset times at 1 AU to the inverse velocities of SEPs
  is used to derive the injection time and path length of the
  first-arriving particles. VDA does not, however, take into account
  that the particles that produce a statistically observable onset at 1~AU
  have scattered in the interplanetary space. We use Monte Carlo test
  particle simulations of energetic protons to study the
  effect of particle scattering on the observable SEP event onset
  above pre-event background, and consequently on VDA results. We find
  that the VDA results are sensitive to the properties of the
  pre-event and event particle spectra as well as SEP injection and
  scattering parameters. In particular, a VDA-obtained path length that
  is close to the nominal Parker spiral length does not imply that the VDA
  injection time is correct. We study the delay to the observed onset
  caused by scattering of the particles and derive a simple estimate
  for the delay time by using the rate of intensity increase at the
  SEP onset as a parameter. We apply the correction to a magnetically
  well-connected SEP event of June 10 2000, and show it to improve
  both the path length and injection time estimates, while also
  increasing the error limits to better reflect the inherent
  uncertainties of VDA.
\end{abstract}

\keywords{Journal approved keywords}

\section{Introduction}\label{sec:introduction}

During solar eruptions, charged particles are accelerated up to
relativistic energies, to form the solar energetic particle (SEP)
population of the cosmic rays observed by in situ instruments at
different locations in the heliosphere. The particles are believed to
be accelerated in flares and CME-driven shock waves
\citep{Reames1999}. However, the relative importance of the flare and
CME processes on the origin of the observed SEP populations is still
under scientific discussion, and opinions differ on how the eruption
phenomena and the SEP production are connected
\cite[e.g.,][]{Cane2010,Gopalswamy2012,Aschwanden2012}.

The difficulty in deducing the particle acceleration scenarios during
solar eruptions stems from the nature of SEP observations. The
propagation of the charged SEPs is affected by the interplanetary
magnetic field. The SEPs are guided by the large-scale Archimedean
spiral structure of the interplanetary magnetic field, the Parker
Spiral. Solar wind is turbulent, and the particles scatter off the
inhomogeneities of the magnetic field
\citep[e.g.,][]{Parker1965}. Thus, the particle propagation is
diffusive rather than direct propagation from the acceleration site to
the in-situ particle detectors, typically at 1~AU from the Sun.  In
order to understand the connections between the components of a solar
eruption and the observed SEP intensities, we must understand the
propagation of SEPs in the interplanetary space, and deconvolve it
from the observations. This has been done for several SEP events
\citep[see, e.g.,][]{Kallenrode1993, Tors96, Laitinen2000, Droge2003,
  AgEa09}, with most recent works introducing cross-field diffusion
into the modelling of SEP events
\citep[e.g.,][]{Zhang2009,Droge2010,He2011,Dresing2012}.

Deconvolving of the interplanetary transport from the in-situ SEP
observations is, however, not simple, and is usually performed only in
case studies. For larger statistical studies, simpler methods to
obtain the injection time of the SEPs are commonly used. The popular
choice is to use the Velocity Dispersion Analysis (VDA), where the
first-observed particles are assumed to have propagated without
scattering \citep[e.g.][]{Lin1981, Reames1985, Torsti1998, KrLi00,
  Tylka2003, Reames2009,Vainio2013_sepserver}. The possible
uncertainties on the arrival times of first particles caused by
interplanetary transport effects are typically not evaluated in these
studies.

The validity of the VDA method has been studied by using numerical SEP
simulations that solve the focused transport equation describing particle
propagation in interplanetary space. By using onset times at 1~AU
obtained from simulated SEP time-intensity profiles at different
energies, it has been shown that for strong scattering conditions the
VDA can result in large errors for the injection times and path
lengths of the particles \citep{KaEa90,Lint04,Saiz05}. These studies,
however, defined the onset time relative to the maximum intensity
(e.g., the time when the intensity reaches 1\% of the maximum
intensity), which is not the common practice when analysing real SEP
events. Unlike in simulated SEP events, real events may have a complex
structure, due to local interplanetary magnetic field structures and
multiple, energy-dependent injection components
\citep[e.g.][]{Laitinen2000}. This may affect the time profile of the
event before the maximum intensity is reached. Thus, to get the best
estimate of the beginning of the injection, the onset of an SEP event
is typically determined by using the moment when the SEP intensity
exceeds the pre-event background by a statistically significant amount
\citep[e.g.,][]{HuHe05}. \citet{LaHu10} analysed simulated SEP events
with VDA by adding a pre-event background to the time-intensity
evolution of the SEP event, and using the background-exceeding time as
the event onset time. They found that the errors in VDA have strong
dependence on the pre-event and event maximum spectra.

In this work, we build on the results of \citet{LaHu10} and discuss
the complexity of the effect of the pre-event background and the role
of scattering in different types of SEP events, and show how different
event types can result in very different VDA results. We study the SEP
time-intensity profile of energetic protons at the time of
the observed onset, and derive a simple estimation for the delay the
particles experience due to the scattering in interplanetary space. We
show that this delay estimate can successfully be used as a correction
to the observed onset times to improve the accuracy of the VDA. We
apply the correction and its error limits to simulated SEP events and
the SEP event of June 10 2000, and show that the method improves the
deduced injection time in both cases.

To clarify the used terminology, we use the term ``\textit{injection
  time}'' to describe the time of release of SEPs at or near the Sun,
``\textit{onset time}'' to describe the time when the SEP intensities
are observed to rise at 1 AU (see discussion below), and
``\textit{launch time}'' as the time when a CME is estimated to lift
off. Furthermore, the symbol $t_i$ refers to the injection time
obtained by VDA, and $t_o$, the onset time obtained from the simulated
events. The time required for a scatter-free particle with velocity
$v$ to propagate distance $s$, i.e., $s/v$, is referred to as
``\textit{scatter-free time}''.

\section{Models}\label{sec:models}

\subsection{Solar particle event modelling}\label{sec:solar-particle-event}

In this work, we study the effect of pre-event background on SEP
injection times as determined by the VDA method, in the presence of
interplanetary scattering and different injection profiles. We solve
the focused transport equation of energetic protons,
\begin{eqnarray}
  \pder{f}{t}&+&\pder{}{z}v\mu
  f +\pder{}{\mu}\frac{v}{2L}\left(1-\mu^2\right)f \nonumber\\
&-&\pder{}{\mu}\left(1-\mu^2\right)\nu\pder{f}{\mu}=Q(z,\mu,t),  \label{eq:foctransp}
\end{eqnarray}
with $f(v,\mu,z,t)$ the particle distribution function, $v$,
  $\mu$, $z$ and $t$ the speed, pitch angle cosine, position along
  field line and time, respectively, $L$ the focusing length, $\nu$
  the scattering frequency, and $Q$ describing the particle source.
  The equation is solved using Monte Carlo simulations, where
  test-particles are propagated and focused along the Parker spiral in
  a frame co-rotating with the Sun, and scattered isotropically in
  pitch angle, with $\nu=v/(2\lambda)$, where $\lambda$ is the
  parallel scattering mean free path \citep[e.g.,][]{Tors96,Koch98}. Adiabatic
  deceleration is taken into account by scattering the particles in
  the co-rotating solar wind frame. Propagation of paricles across the
  Parker spiral direction is not considered.

We use the initial energy range from 1~to 120~MeV with a power law
spectrum $\propto E^{-3}$. This power law is used for all
  runs, to reduce the parameter space. The particles are injected at
0.01~AU, at time $t=0$, with a reflecting boundary at the Sun,
  and followed for 48~hours. The Parker spiral is parametrised by a constant
solar wind velocity of 400~km/s and solar rotation period of
  25.35 days, resulting in Parker spiral length of 1.17~AU. We use
constant radial mean free path, with values $\lambda_r=0.3$~AU at 1GV
rigidity to represent moderate scattering conditions, and 1~AU at 1~GV
rigidity to represent weak scattering conditions
\citep[e.g.][]{Palmer82}. The mean free path is taken to depend on the
particle rigidity, as $R^{1/3}$, consistent with quasilinear
  theory for Kolmogorov slab spectrum \citep{Jokipii1966} and
  observations \citep[e.g.][]{Droge2000ApJ}.

The Monte Carlo simulations are used to obtain a response, as observed
at 1~AU, for an impulsive injection of protons, at 20 logarithmically
spaced energy channels between 1~and 100~MeV. The upper limit
  of the channels is chosen to be 100~MeV rather than 120~MeV, because
  particles experience adiabatic deceleration. In
order to mimic more realistic particle release scenarios, we convolve
the impulsive responses with two injection profiles at the Sun: a
\emph{fast injection} model, with 30 minutes of linear increase of
injection strength, followed by 570 minutes of linear decay; and a
\emph{slow injection} model, with 570 minutes of linear increase,
followed with 30 minutes of linear decay. These
profiles were selected because CME-related SEP acceleration has been
estimated to be most efficient when the CME is at 5--15~R$_\odot$
  \citep[e.g.][]{Kahler1994}. A CME with velocity 2000~km/s reaches
  5~R$_\odot$ in half an hour, corresponding to our fast injection
  model. The slow injection model accounts for slower CMEs, higher
  maximum injection heights, poor connection between the spacecraft
  and acceleration region and possible cross-field transport of SEPs,
  that may cause a gradual injection of particles.

\begin{figure}
\includegraphics[width=\columnwidth]{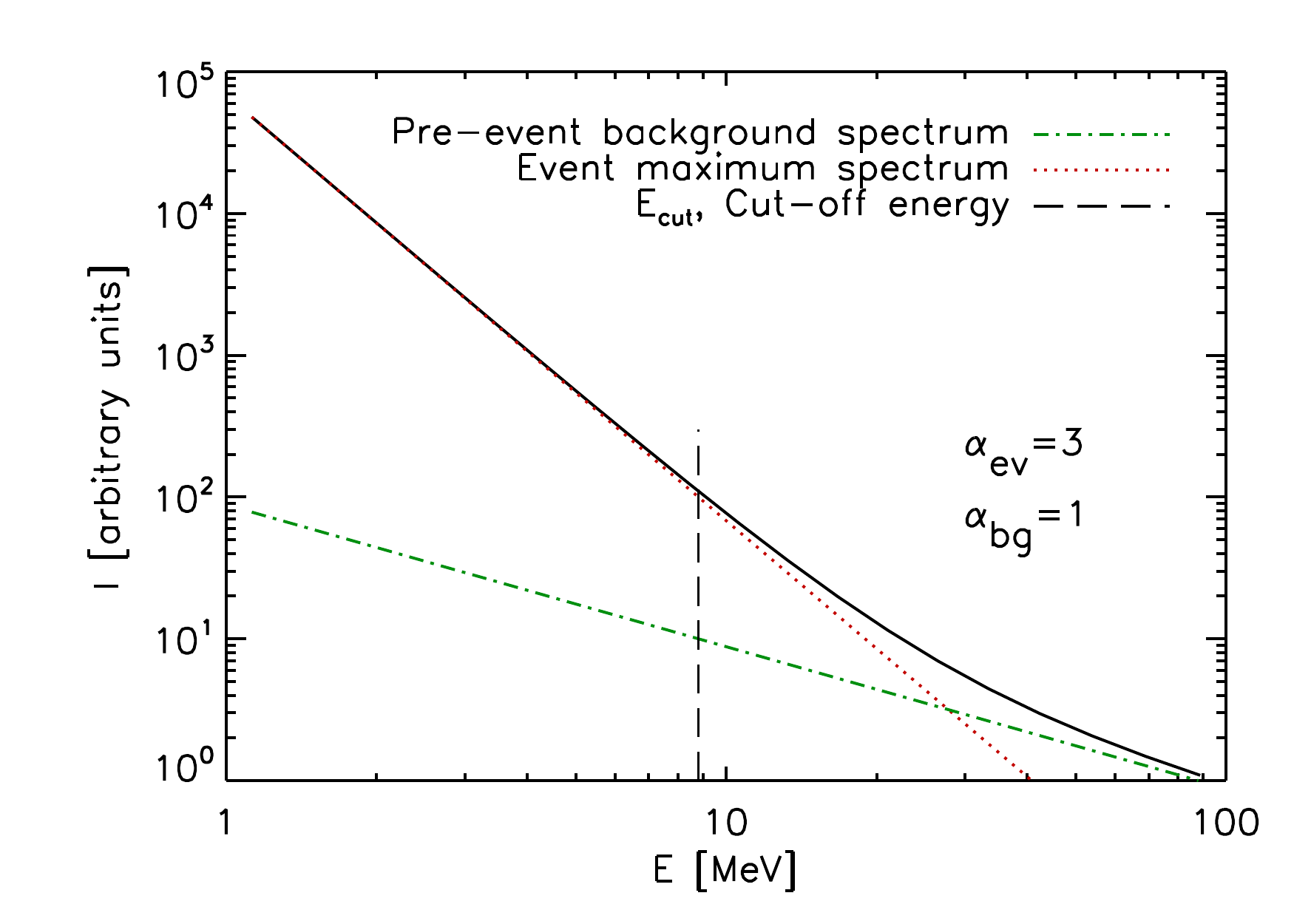} 

\caption{An example of the event maximum spectrum (red dotted
  curve),the pre-event background spectrum (green dash-dotted curve),
  with the total intensity spectrum at event maximum given by solid
  black curve. The dashed vertical black line depicts the energy
  $E_{cut}$, where the event maximum intensity is an order of
  magnitude times the background intensity. The event maximum spectrum
  power law index is~3, and the background power law
  index~1.\label{fig:spectra}}
\end{figure}

We define our study parameters using SEP intensities that can
  be observed directly at 1~AU. We scale the convolved time-intensity
  profiles of simulated SEPs at 1~AU so that the spectrum calculated
  from the maxima of the scaled intensities at each channel forms a
  power law,
\begin{equation}
  I_{ev}  =  I_{0ev}\left(\frac{E}{E_{0}}\right)^{-\alpha_{ev}}
\end{equation}
where $I_{0ev}$ is the SEP event maximum intensity at energy
$E_{0}=88$~MeV, and $\alpha_{ev}$ the power law index of the spectrum.
The resulting event time-intensity profiles are overlaid on a constant
pre-event background, which also follows a power law energy spectrum, 
\begin{equation}
  I_{bg}=I_{0bg}\left(\frac{E}{E_{0}}\right)^{-\alpha_{bg}}
\end{equation}
where  $I_{0bg}$ is the background intensity at $E_{0}$ and
$\alpha_{bg}$ the background power law index. An example of the
spectra is shown in Fig.~\ref{fig:spectra}.

\subsection{Velocity dispersion analysis}\label{sec:veloc-disp-analys}

The Velocity Dispersion Analysis is based on the assumptions that the
first particles observed at a given distance from the Sun have been
released simultaneously, propagate the same path length, and
experience no scattering or energy changes. Under these conditions,
the arrival time, $t_{o}$, of the particles to the observer at
distance $s$ along the magnetic field line is given by
\begin{equation}
t_{o}(v)=t_{i}+\frac{s}{v},\label{eq:VDA}
\end{equation}
where $t_{i}$ is the particles' injection time at the Sun, $s$ the
travelled distance and $v$ the particle velocity. Thus, knowing the
observed onset times at 1~AU, and the velocities of the particles,
a simple linear fitting of this data according to the Eq.~(\ref{eq:VDA})
gives the particles' injection time at the Sun, and the path length
traveled by the particles. In the simulated events, the injection of
the particles starts at $t=0$, thus a successful VDA fit would give
$t_i=0$ and $s=1.17$~AU.

While all of the VDA assumptions can be questioned, in this study we
concentrate only on the effect of the scattering on the derived
injection time and path length. In particular, we will study the
common practice of defining SEP onset as the moment the intensity is
discernible from the pre-event background, and how this practice
affects the VDA results.

The SEP event onset time, observed by particle detectors, is often
difficult to determine due to low counting rate of particles of the
ambient energetic particle population before and at the very beginning
of the event. To determine the onset time, typically a threshold of
one or several standard deviations above the pre-event background is
used to define the onset in a statistically significant way \citep[see
also][for an alternative method]{HuHe05}. This results in a delay for
the observed onset. In order to mimic this effect, we define the onset
to be observed when the intensity at 1~AU rises 10\% above the
pre-event background.

\section{Results}\label{sec:results}

\subsection{Parametric study}

\begin{figure}
\includegraphics[width=\columnwidth]{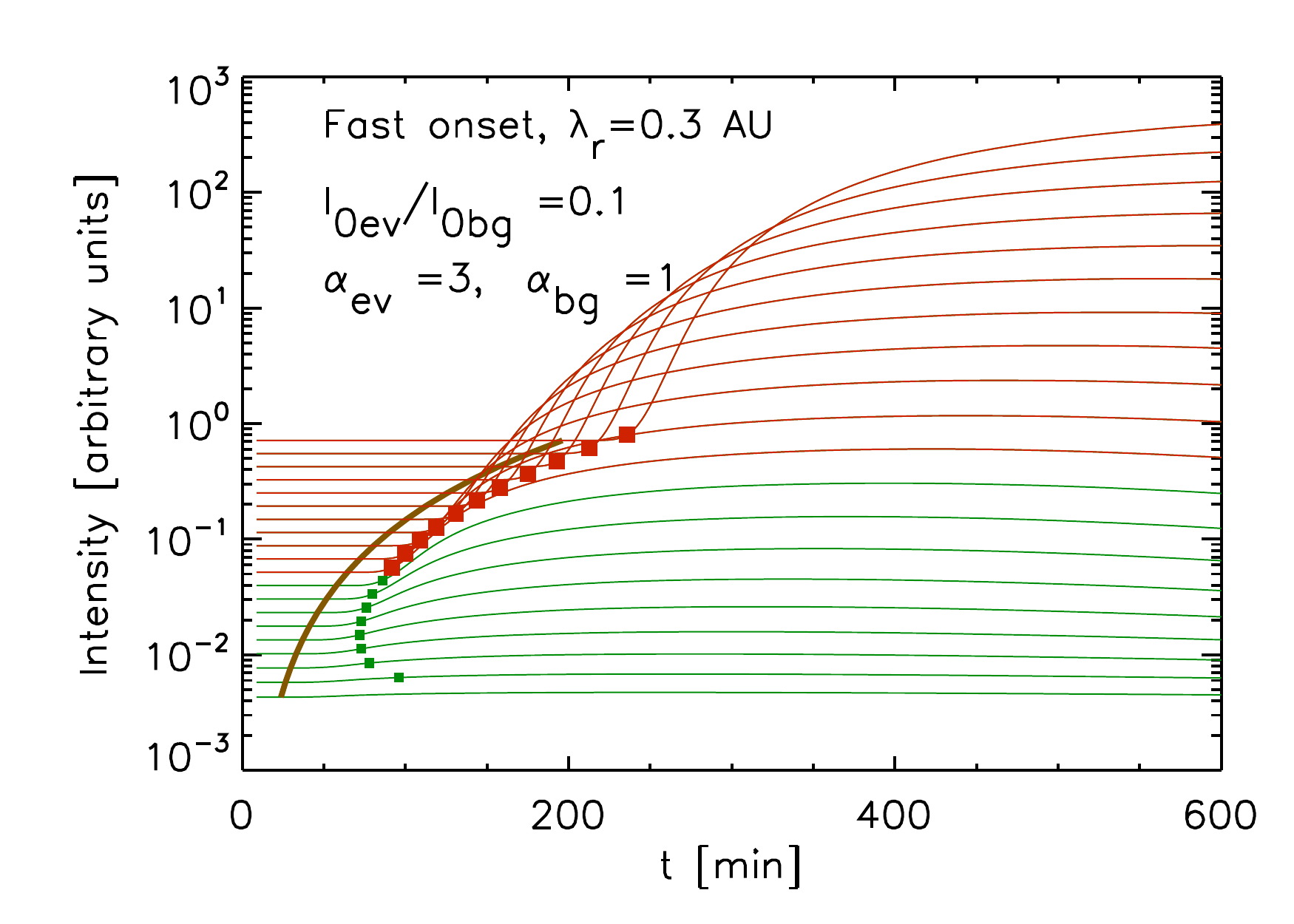}

\caption{An example of a proton event with energies
    1--100~MeV overlaid on a pre-event background. The lowest-energy
  intensities are on the top of the figure, the highest on the bottom.
  The onset times, as defined in the text, are marked by squares.  The
  red curves and the large red squares represent the channels where
  the event maximum intensity is over 10 times the bacground, the
  green curves and the small green squares represent the
  smaller-intensity energies. The thick brown curve represents the
  scatter-free arrival time of particles.\label{fig:eventexample}}
\end{figure}

\begin{figure}
  \includegraphics[width=\columnwidth]{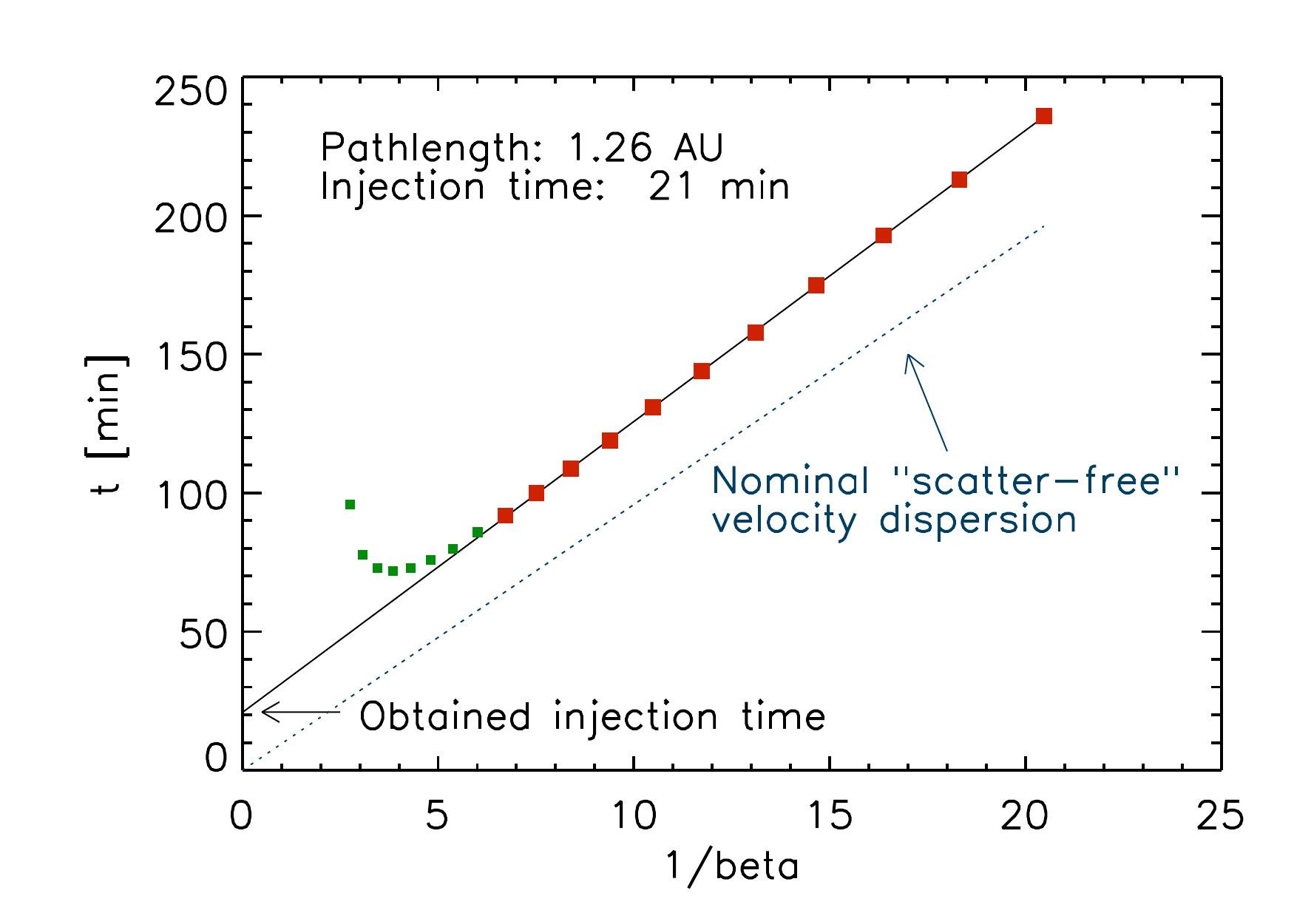}

\caption{An
    example of the linear fit of the observed onset times and the
    inverse velocities. The squares depict the observed onset times,
    with the color scheme the same as in
    Fig.~\ref{fig:eventexample}. The thin black curve depicts the
    resulting linear fit to the energy channels with event maximum
    intensity over 10 times the pre-event background.  The obtained
    parameters of this fit are given on the top-left corner of the
    figure. The dashed blue line depicts the velocity dispersion for
    particles that are observed after time $t=1.17$~AU$/v$ from their
    injection.\label{fig:binvexample}}
\end{figure}

We show an example of an analysed event in
Figs.~\ref{fig:eventexample} and~\ref{fig:binvexample}. In
Fig.~\ref{fig:eventexample}, we show the time evolution of a simulated
event, with maximum intensity spectrum $\propto E^{-3}$, taking place
when the pre-event background spectrum is proportional to $E^{-1}$,
with the event intensity an order of magnitude below the background
intensity at energy 88~MeV, and $\lambda_r=0.3$~AU, and fast
  injection. The onset times of the intensity increases are shown by
symbols, with the brown thick curve crossing the time-intensity curves
on the left showing the theoretical time of arrival of particles in
the case of particles following the assumed 1.17~AU Parker Spiral
field without scattering. As can be seen, the intensity increase
begins significantly later than the scatter-free time at all energy
channels.

The effect of the delay on the VDA can be seen in
Fig.~\ref{fig:binvexample}, where we show the onset time plotted
against $1/\beta$ (the red and green symbols). As can be seen, the
onset times are clearly delayed from the velocity dispersion pattern
for particles that are observed immediately after scatter-free
propagation to 1.17~AU (dashed blue line). In addition, not all
observed onsets follow the linear velocity dispersion pattern. In our
example event, at large energies the arrival time dependence on
$1/\beta$ is inverted, with onsets at higher energeies observed later
than at lower energies, due to the pre-event background. Such energy
channels clearly must be excluded from the velocity dispersion fit. In
order to achieve this, we have excluded energy channels where
$I_{ev}(E)/I_{bg}(E)<10$ (the green curves and small symbols in
Figs.~\ref{fig:eventexample} and~\ref{fig:binvexample}).

However, even after removing the non-linearly behaving energy channels
from the velocity dispersion fit, the pre-event background and
scattering still have a substantial effect on the determination of the
SEP injection time. The observed onset time at different energy
channels is delayed by 20--40 minutes from the scatter-free arrival
time. The VDA-fitted injection time for this event is 21 minutes later
than the actual release time of particles at the Sun. The path length,
1.26~AU, is clearly longer than the nominal 1.17~AU Parker spiral
length in our model. The increased path length is due to the
low-energy (high $1/\beta$) particles having longer delay relative to
the scatter-free arrival time (blue dashed line in
Fig.~\ref{fig:binvexample}), as compared to the higher-energy
particles.

\begin{figure*}
  \includegraphics[width=\textwidth]{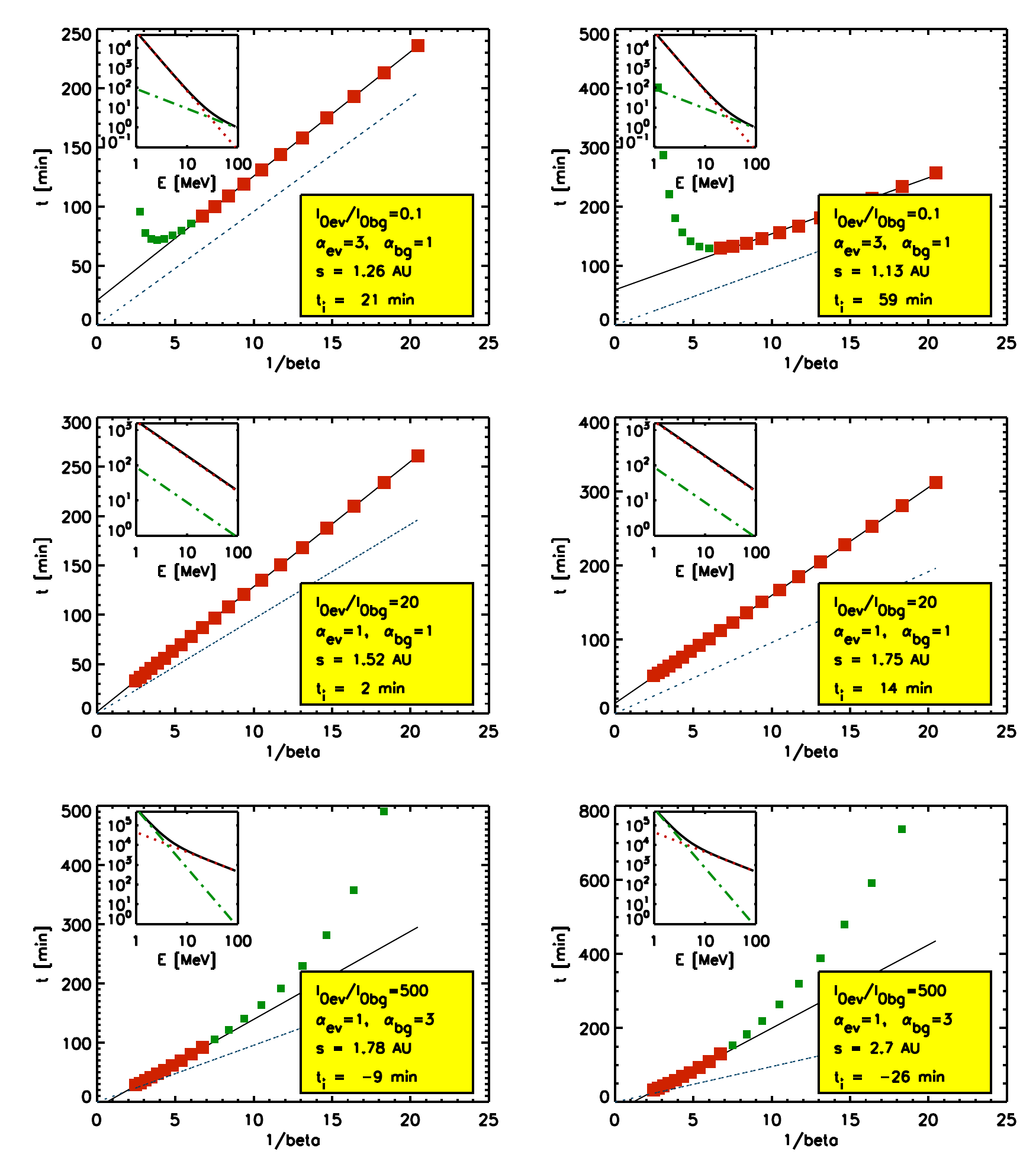}

\caption{Onset times fitted
    for simulated events with different event maximum and background
    spectra, and injection profiles. The left column presents the
    results of the fast injection model, and the right the slow
    injection model,
    with $\lambda_r=0.3$~AU. The symbols represent the obtained onset
    times at different energies, with the large {[}small{]} symbols
    for channels where the intensity at event maximum is more
    {[}less{]} than an order of magnitude above pre-event
    background. The solid curve shows the $1/\beta$-fit to the onsets,
    and the dashed curve the fit expected for scatter-free
    particles. The inset in the top left of the panels shows the event
    maximum spectrum (dashed curve), background spectrum (dot-dashed
    curve) and the total spectrum (solid curve).\label{fig:fits}}
\end{figure*}

In Fig.~\ref{fig:fits}, we show the VDA fits for $\lambda_r=0.3$~AU
for the fast (left column) and slow (right column) injection model,
and different types of events. In the top panels, we show an SEP event
with the maximum spectrum softer than pre-event background
spectrum. This type of event may be observed when the
pre-event background is not affected by preceding SEP events \citep[e.g.][]{Valtonen2001}. In such
an event, the intensities at higher energies are masked more
efficiently by the pre-event background than at lower energies. As can
be seen, the VDA path lengths are typically close to or slightly
longer than the nominal Parker spiral length, with the obtained
injection time considerably later than the actual solar injection
time.

In the event shown in the middle panels, the pre-event and event
maximum spectral indices are equal. This can take place for example
when a soft-spectrum SEP event follows a harder-spectrum SEP
event. The delay with respect to the scatter-free arrival time is
significantly larger at low energies. This results in a significantly
longer path length, of 1.52 AU in the case of fast injection model
(left middle panel), and 1.75~AU for the slow injection model (right
middle panel). The long path length, however, compensates partly the
delay caused by scattering, resulting in only 2-minute error for the
injection time determination with the fast injection model.

The bottom panels of Fig.~\ref{fig:fits} depict an event where a hard
SEP event takes place on a soft pre-event background. This typically
takes place when a previous SEP event is still decaying at the time of
a new SEP event. In this case, the lower energies are efficiently
masked by the pre-event background, and the velocity dispersion fit
results in a very long path length. As can be seen in both the fast
and slow injection models (bottom left and right panels,
respectively), the obtained injection time can precede the real
injection time in such cases.

Fig.~\ref{fig:fits} shows that also the injection profile of the
particles has a significant effect on the VDA results, with the
comparison of fast and slow injection profiles on the left and right
columns, respectively. As can be seen, the effect of the pre-event and
event maximum spectral shapes on the velocity dispersion pattern is
similar for both injection models. However, the delay times for
individual energy channels are longer in the slow injection case, and
the resulting error in injection time is also larger.

As shown in Fig.~\ref{fig:fits}, the assumption of scatter-free
propagation can result in significant systematic errors for the SEP
injection time and the traversed path length. This error is caused by
the delay of particles due to interplanetary scattering, which should
be taken into account in the velocity dispersion analysis as
\begin{equation}
t_{o}(v)=t_{i}+\frac{s}{v}+t_d(v,\lambda,t_o),\label{eq:VDA_td}
\end{equation}
The estimation of the delay time $t_d$ is not trivial, as it depends
on scattering conditions, injection profile and the level of the
pre-event background relative to the SEP event intensity. Also the
estimation of the resulting error in VDA is not straigthforward: as
shown in the middle panels of Fig.~\ref{fig:fits}, an error in the
path length can compensate the error in the injection time
determination. Such a complicated relation between the observed
velocity dispersion and the injection is difficult to analyse, and may
skew the results of large statistical SEP studies, where fitting of
the transport of SEPs is not feasible.

\subsection{Analysis of the delay time in VDA}

The rise time of SEP intensities above the pre-event
background is not the only observable of an SEP event onset. As shown
in Fig.~\ref{fig:eventexample}, the intensities rise initially roughly
exponentially, with different rise rates at different energies and
intensities. This gradual rise, as opposed to immediate rise to the
maximum intensity, is a result of the particle scattering in
interplanetary space. In the following, we study the rate of intensity
increase at the time of SEP event onset.

For a diffusion process with scattering mean free path $\lambda$, the
intensity of particles with velocity $v$ at distance $r$ and time
$t$ is given as
\begin{equation}
  I_{\mathrm{diff}}(r,t)\propto \frac{1}{\left(4\lambda v t/3\right)^{3/2}}
  \exp\left\{-\frac{3 r^2}{4 \lambda v t}\right\}.
  \label{eq:diffusion}
\end{equation}
From this, we can obtain the timescale of the intensity increase as
\begin{equation}
\tau_I^{-1}\equiv\frac{1}{I_{\mathrm{diff}}}\frac{\mathrm{d}
  I_{\mathrm{diff}}}{\mathrm{d}t}=\frac{3}{2}\frac{1}{t}\left(\frac{1}{2}\frac{r}{\lambda}\frac{r}{v\,
    t}-1\right)\label{eq:diffderiv}
\end{equation}
The diffusion approach is valid only for times $t\gg\lambda/3v$ and
$t\gg r/v$. Indeed, the intensity maximum takes place at
$t_{max}=r/(2\lambda)\; r/v$, which for $\lambda> r/2$ is before the
particles with velocity $v$ can arrive scatter-free to distance $r$,
clearly an unphysical result. However, as there are no analytic
descriptions for the initial phase for diffusively spreading
particles, we will use diffusion as a starting point of our study.

We are only interested in the initial increase of the SEP event, and
use only times when the intensity is less than $I_{ev}(E)/10$. Thus we
only use the first term in Eq.~(\ref{eq:diffderiv}). Rewriting
$\tau_I$ using the diffusion timescale, $\tau_D=3r^2/(2\lambda v)$, we
arrive with a simple scaling between the time from SEP injection at
the sun, $t$, and the intensity increase timescale $\tau_I$ as
\begin{equation}
\frac{1}{2}\tau_I \tau_D=t^2.\label{eq:taui_taud_t}
\end{equation}

\begin{figure}
  \includegraphics[width=\columnwidth]{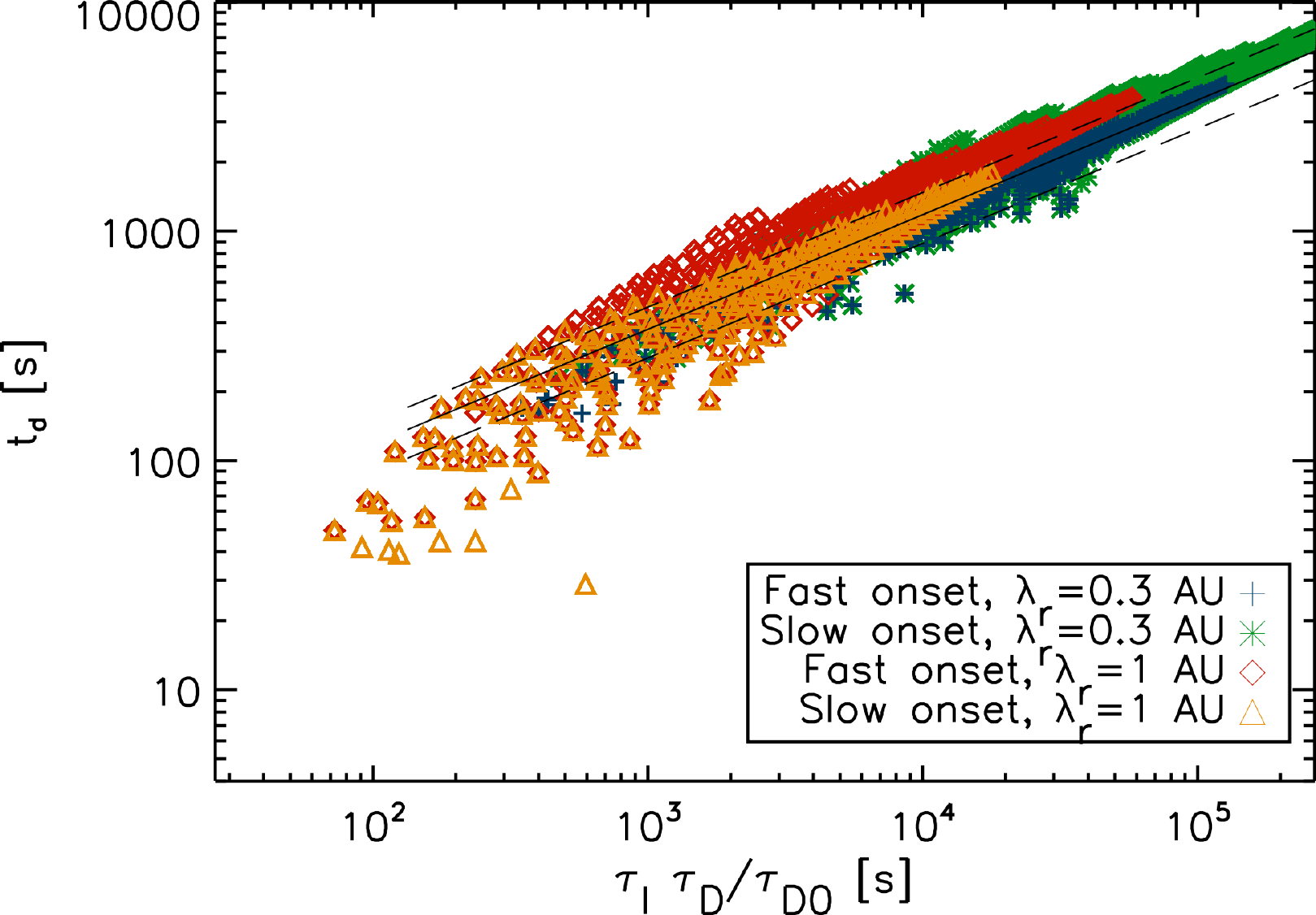}

\caption{Scatter plot of $t_d=t_{i}-s/v$ against the scaled intensity
    increase scale time, for simulations with different mean free
    paths and injection profiles. The solid curve represents the best
    fit of the Eq.~(\ref{eq:tsys_theor_noncorr}) to the simulation
    data, with the dashed curves depicting the error limits of the
    fit.\label{fig:ascensionfit}}
\end{figure}

To study this scaling, we simulated a large number of SEP events with
the ratio $I_{0ev}/I_{0bg}$ ranging from $10^{-4}$ to $10^{4}$, and
the difference of the event maximum and pre-event background spectral
indices from~-3 to~3, both fast and slow injection models, and
moderate and weak scattering conditions. The intensity increase
timescale was obtained from the simulated events by using intensities
at two consecutive times, with
\begin{equation}
\tau_I^{-1}=\frac{\ln\left(I_{ev}(t_{2})/I_{ev}(t_{1})\right)}{t_{2}-t_{1}}.\label{eq:tau_I}
\end{equation}
The scaling in Eq.~(\ref{eq:taui_taud_t}) is obtained by multiplying
the intensity increase timescale by $\tau_D/\tau_{D0}$.  As the radial
mean free path $\lambda_r$ is constant in our model, we use the
$\lambda_r$ and $r$=1~AU in the diffusion timescale. The reference
value, $\tau_{D0}$ is calculated using $\lambda_r=1$~AU and $v=c$.

\begin{figure*}
  \includegraphics[width=\textwidth]{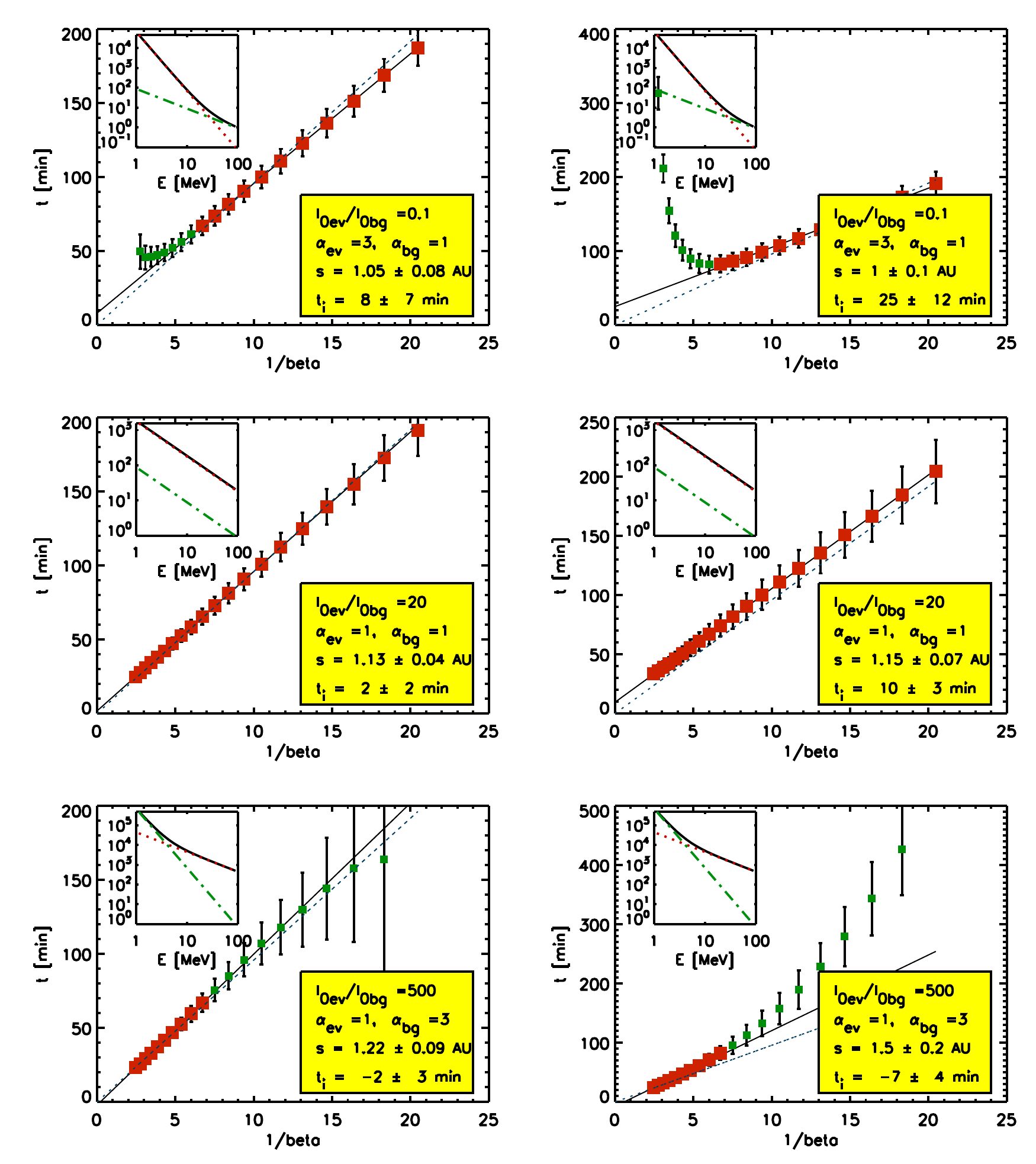}

\caption{Onset times
    fitted for the same simulated events shown in Fig.~\ref{fig:fits},
    with the onset times corrected using
    Eq.~(\ref{eq:tsys_theor_noncorr}). The injection profiles,
    parameters, symbols and curves are as in
    Fig.~\ref{fig:fits}.\label{fig:fits_fix}}
\end{figure*}

We found that comparing $\tau_I$ with the onset time did not result in
the expected power law behaviour given by Eq.~(\ref{eq:taui_taud_t}),
due to the inability of the diffusion description to describe the
early time evolution of SEP propagation. However, as shown in
Fig.~\ref{fig:ascensionfit}, the power law behaviour is retained when
plotting $\tau_I\tau_D/\tau_{D0}$ as a function of delay time $t_d$,
for several orders of magnitude, for different particle energies,
scattering conditions and injection profiles. The delay time is a
reminescent of the concept of signal speed in modelling the particle
propagation with the Telegraph equation \citep{Fisk1969}. However, as
shown by \citet{Effenberger2014}, the Telegraph equation models the
initial phase of an SEP event poorly, and is not hence addressed
further in this study.

As seen in Fig.~\ref{fig:ascensionfit}, the power law suggested by
Eq.~(\ref{eq:taui_taud_t}) is not exact. Thus, to estimate the delay
time for using it as a correction for VDA, as in
Eq.~(\ref{eq:VDA_td}), we fitted the data in
Fig.~\ref{fig:ascensionfit} to
\begin{equation}
t_d=a\,\sqrt{\frac{\tau_I \tau_D}{2}}\label{eq:tsys_theor_noncorr}
\end{equation}
which gave $a=0.61$, 
with standard deviation $\sigma_a=0.15$. 
This estimate can be used further in correcting for the effects of
interplanetary transport on VDA. At larger values of $\tau_I$,
  the diffusive profile is approached and values of $a$ up to unity could be used.

The data can also be fitted as a power law, resulting in power
  law index of 0.56, instead of the form suggested by
  Eq.~(\ref{eq:taui_taud_t}). However, we consider the fit to
  Eq.~(\ref{eq:tsys_theor_noncorr}) a better choice, as it is based on
  the physics of particle propagation at the time-asymptotic limit,
  and as such is more likely applicable to other energy ranges and
  particle species.

We have applied the correction to the observed onset times given by
Eq.~(\ref{eq:tsys_theor_noncorr}) to the six simulated event examples
shown in Fig.~\ref{fig:fits}, and present the resulting, corrected VDA
fits in Fig.~\ref{fig:fits_fix}. The delay time $t_d$ given by
Eq.~(\ref{eq:tsys_theor_noncorr}) has been subtracted from the onset
times, $t_o$, and the error limits as defined by $\sigma_a$ have been
used in the fitting to obtain error limits for the injection time and
the path length. As can be seen, the correction improves both the
injection time and the path length estimates considerably. The
injection times are correct within the error limits for the fast
injection profile (left column), with the path lengths also better
reflecting the Parker Spiral length of 1.17~AU.

When analysing real SEP events, the limited counting statistics of the
particle detectors cause an uncertainty factor for the delay time
through the dependence of $\tau_I$ on intensity. We assume that the
pre-event background intensity can be evaluated from sufficiently long time
period so that its statistical error is insignificant compared to
those of intensities at times $t_1$ and $t_2$ in
Eq.~(\ref{eq:tau_I}). With this assumption, the statistical error of
$\tau_I$ can be evaluated as
\[
\frac{\Delta\tau_I^{2}}{\tau_I^2}\approx\frac{1}{\left[\ln\left(I_2/I_1\right)\right]^2}\left(\frac{1}{N_{1}}+\frac{1}{N_{2}}\right),
\]
where the variables $N_{1}$ and $N_{2}$ represent the number of
particles detected by the instrument at the two times used to
calculate $\tau_I$. Using this approximation, we have
\begin{equation}
\frac{\Delta
  t_d}{t_d}\approx\sqrt{\left(\frac{\sigma_{a}}{a}\right)^{2}+\frac{1}{4}\frac{N_{1}+N_{2}}{N_{1}N_{2}\left[\ln\left(I_{2}/I_{1}\right)\right]^{2}}}.\label{eq:error}
\end{equation}
If the relative error from the background determination is
significant, it should be included in the error analysis.

In addition to the finite counting statistics, there is still one
significant source of uncertainty in
Eq.~(\ref{eq:tsys_theor_noncorr}). The diffusion timescale, $\tau_D$,
depends on the scattering mean free path of the particles,
$\lambda$. While $\lambda$ depends on the amplitude of interplanetary
turbulence, it is difficult to estimate, and is typically obtained as
a side product of SEP transport fitting. According to several studies
\citep[e.g., ][]{Palmer82}, the mean free path varies from 0.08 to~0.3
AU, with some recently analyzed events showing significantly longer
mean free paths \citep[e.g.,][]{Torsti2004}. There is no reason to
expect the mean free paths to be normally distributed, thus we
evaluate its effect on the delay time estimation by using extreme mean
free path values, e.g., 0.1~and 1.0~AU, and obtaining the smallest and
largest parameter values (taking also the statistical errors into
consideration).

\subsection{June 10 2000 SEP event}

\begin{figure}
\includegraphics[width=\columnwidth]{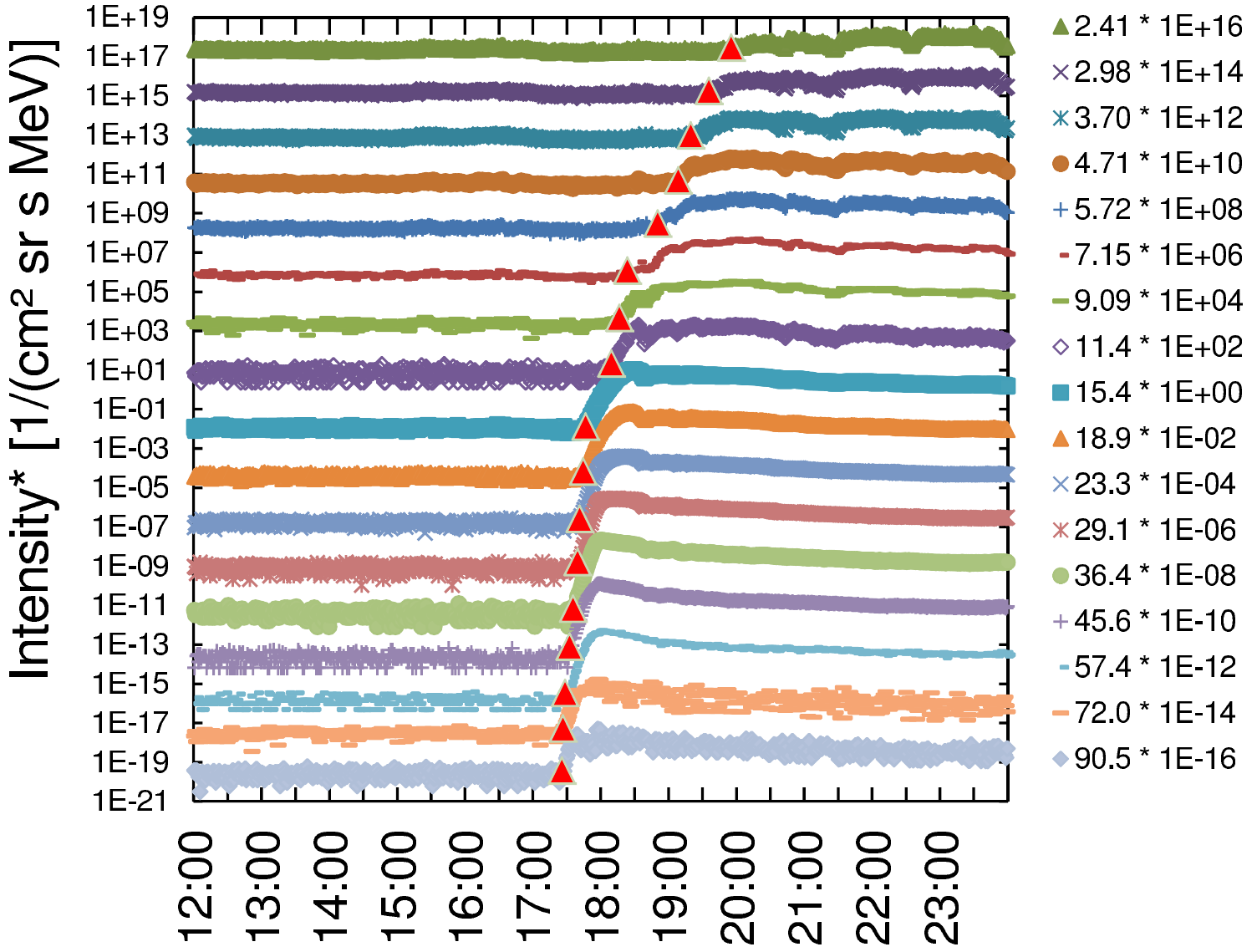}

\caption{Proton intensities at energies 2.41--90.5~MeV on June 10, 2000, as observed by the ERNE
  instrument onboard SOHO spacecraft. The onset times, as determined
  by method described in \citet{HuHe05}, are shown with red
  triangles. The intensities in each consecutive energy channel are
  multiplied by a factor of 100 to better distinguish onset
  times.\label{fig:event_int} }
\end{figure}

\begin{figure}
\includegraphics[width=\columnwidth]{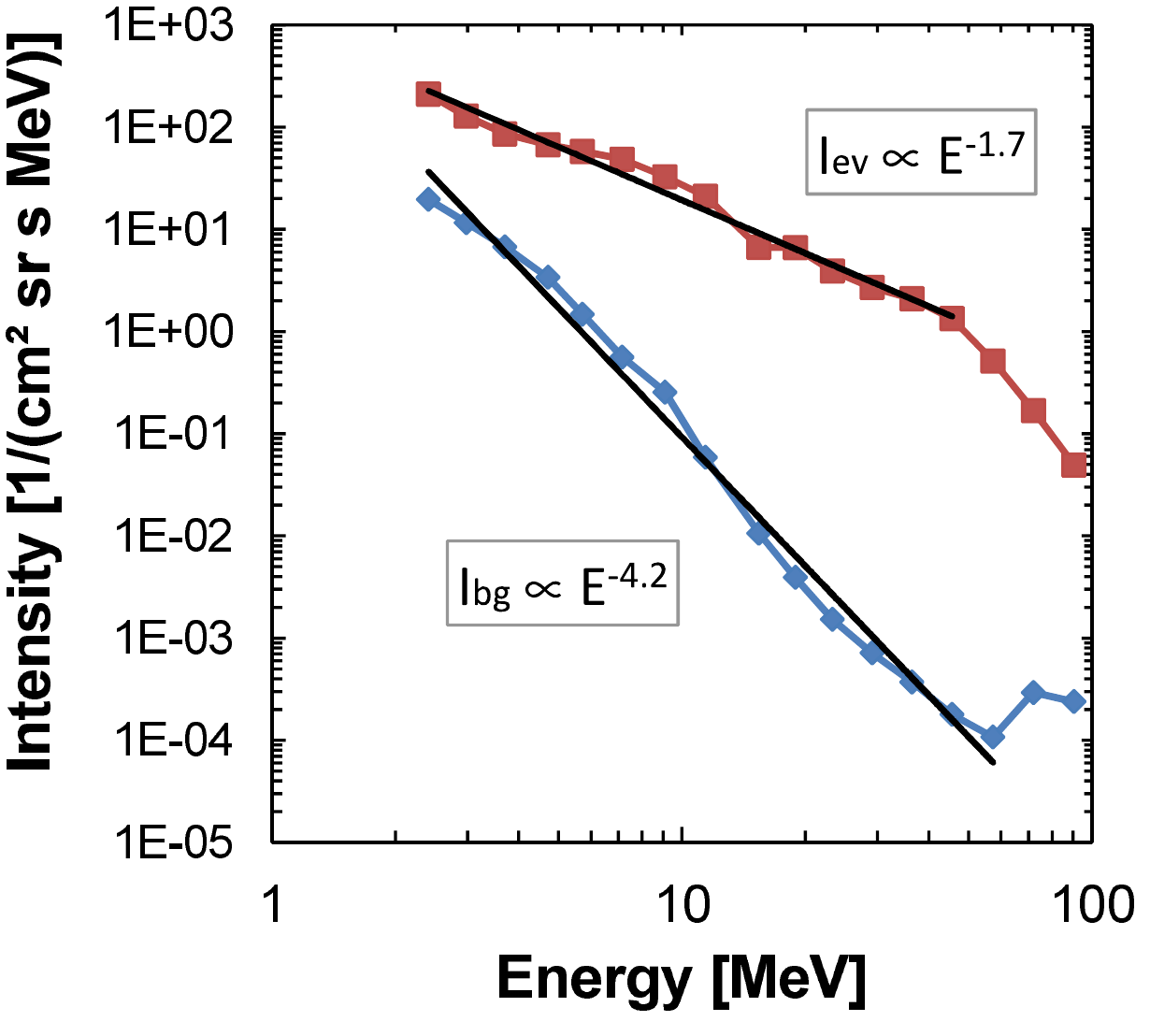}

\caption{Event maximum (red curve and squares) and pre-event
  background (blue curve and diamonds) spectra on June 10, 2000, with
  respective power law fits. \label{fig:event_spectra}}
\end{figure}

In order to study the effects of the delay due to scattering, and the
ability of the correction derived above to improve the VDA
results, we apply the correction to an observed SEP event. In
Figs.~\ref{fig:event_int} and~\ref{fig:event_spectra}, we show the
time-intensity profiles and spectra of the SEP event of June 10, 2000, as
observed by ERNE instrument onboard SOHO spacecraft
\citep{Torsti1995}. The event takes place during a decay phase of an
earlier SEP event. For this reason, the pre-event spectrum is very
soft compared to the event maximum spectral index, as shown in
Fig.~\ref{fig:event_spectra}. Thus, this event corresponds to the
simulated events in the bottom row of Fig.~\ref{fig:fits}.

The SEP event coincides with an M5.2 solar flare and a western
  halo CME. The flare, located at heliographic coordinates N22 W38,
  started at 16:40~UT and reached its maximum at 17:02~UT. The CME was
  launched at 16:45~UT, as given by a linear extrapolation of
  SOHO/LASCO observations to solar surface in the CDAW SOHO/LASCO CME
  list \citep{Gopa2009}. The western location of the eruption implies
a good magnetic connection for the SEPs along the Parker spiral to the
near-Earth spacecraft, with negligible cross-field propagation effects
on the first-observed particles.

\begin{figure}
\includegraphics[width=\columnwidth]{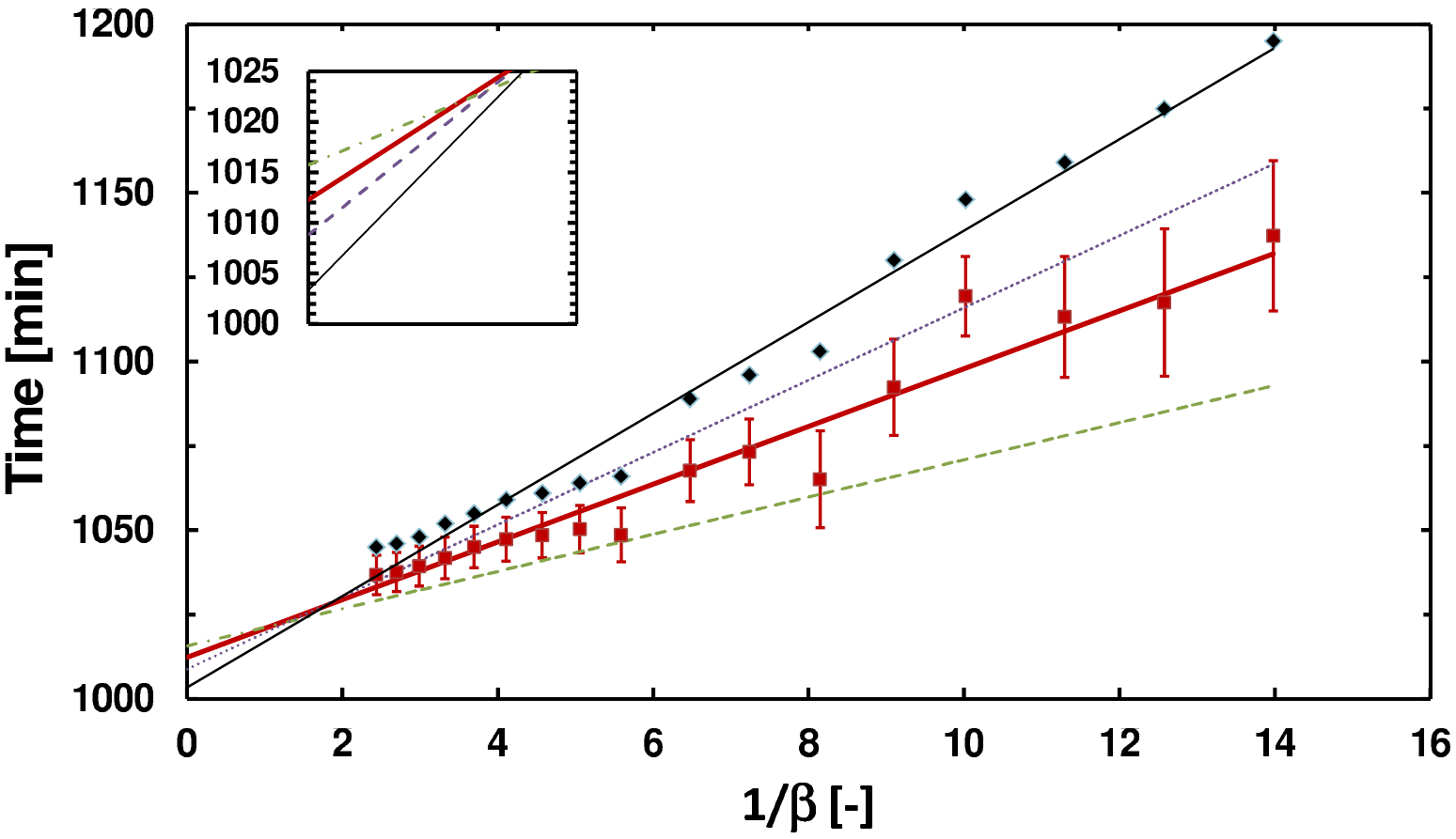}
\caption{Onset time and velocity dispersion fit correction illustration for
June 10th, 2000 event. Observed onset times are represented by black
diamonds, and thin black line is the corresponding least-squares linear
fit ($s=1.63\pm0.06$~AU; $t_{0}=16:43\pm$4~min). Red squares represent
the corrected onset times using reference mean free path 0.3~AU.
See the text for details of the error bars. Blue dotted, thick red,
and green dash-dotted lines correspond to weighted least-squares linear
fits for corrected onset times of reference mean free paths 1.0,~0.3
and 0.1~AU, respectively. These three corrected velocity dispersion
fits can be incorporated to corrected values of $s=1.03_{-0.28}^{+0.34}$~AU,
and $t_{0}=16:52_{-8\,\mbox{min}}^{+7\,\mbox{min}}$. Eight minutes
need to be added to the times to make them comparable to electromagnetic
observations.\label{fig:event} }
\end{figure}

We used the VDA method to determine the solar injection time and path
length of the 2.41--90.5~MeV protons in this event. Using uncorrected
onset times, VDA gave for the solar injection 16:43$\pm$4~minutes
solar time. Taking into account the 8 minutes light needs to travel
from the Sun to Earth, the particles are injected 11 minutes before
the flare maximum time, with the CME estimated to be low in the
corona. The path length of the particles obtained from uncorrected
onset times is very long, $s=1.63\pm0.06$~AU. This is consistent with
the simulated events (bottom row of Fig.~\ref{fig:fits}): an event
with maximum spectrum harder than the pre-event spectrum will show a
very long path length.

We then applied the correction, as given by
Eq.~(\ref{eq:tsys_theor_noncorr}), to the onset times. We show the
effect of the correction in Fig.~\ref{fig:event}, where the black
diamonds and the black line correspond to the uncorrected onset times
and the corresponding VDA fit. The red squares show the corrected
onset times, with the associated errorbars, for mean free path
$\lambda=0.3$~AU. We calculated the corrected VDA results using three
different mean free paths, depicted with the blue dotted, solid red
and green dash-dotted curves in Fig.~\ref{fig:event}. As can be seen
in the inset of the Figure, all three fits give later time of SEP
injection, as compared to the VDA fit without time correction.

Using the errors given by Eq.~(\ref{eq:error}) in the velocity
dispersion fit, we obtain error limits for each of the fits with
different $\lambda$ values. Using injection time, path length and the
fitting error values we find the corrected injection time as
16:52$_{-8\,\mbox{min}}^{+7\,\mbox{min}}+8\,\mbox{min}$, with path
length $s=1.03_{-0.28}^{+0.34}$~AU. Thus, the path length after the
correction is consistent with the expected Parker spiral length. The
solar injection time, of 16:59~UT corrected for electromagnetic
observations, coincides well with the maximum time of the X-ray
flare. The CME was observed at 17:08~UT at 2.76~R$_\odot$. Thus,
within the error limits of the injection time, both the flare and CME
observations are consistent with potential energetic particle
production.

\section{Discussion}\label{sec:discussion}

As this study shows, the results of VDA fitting should be used
carefully. SEPs scatter in the interplanetary medium, and while the
first particles related to an event may indeed be scatter-free, their
intensity is likely too low for statistically reliable observation. It
should be noted that having a very low level of pre-event
counting rates, such as in the case of heavy elements, does not imply that
the first observed SEP event particles are scatter-free. The
instrument's detection threshold has a similar effect to the observed
event onset as the pre-event background intensity level.

It is important to notice that a reasonable path length does not imply
a good estimate for the solar injection time. This can be seen in the
fits presented in Fig.~\ref{fig:fits}: If the path length is nominal,
as in the top panels, all of the fitted energies will have almost
equal delay time $t_d$, which results in a large error for the
injection time. If, on the other hand, the path length is long, the
delay time is shorter at higher energies (smaller $1/\beta$), and the
VDA-fitted curve converges towards the scatter-free VDA pattern
(middle panels of Fig.~\ref{fig:fits}) at the limit of $1/\beta=0$,
and the resulting error in the VDA injection time is small. This can
take place in particular when the spectral indices of the pre-event
background and event maximum spectra are similar. In this case
  the ratio of maximum and background intensities is independent of
  energy, and the VDA result is similar to the method used by
  \citet{Lint04} and \citet{Saiz05}, who determine the onset time as the
  time when a fixed fraction of maximum intensity at the energy
  channel is reached.

It should be remembered that also factors other than the parallel
scattering may influence the particle propagation in the
interplanetary space. Recent multi-spacecraft analyses of SEP events
suggest that SEP events have a wide extent in heliographic longitudes
\citep[][]{Dresing2012,Dresing2014,Wiedenbeck2013,Richardson2014}. While
the large longitudinal spread of particles may be caused by processes
low in the corona \citep[see, e.g.,][for discussion]{Wiedenbeck2013},
the observed longitudinal dependence of anisotropy suggests that
significant interplanetary cross-field transport is taking place
\citep{Dresing2014}. While cross-field diffusion has traditionally
been suggested to transport SEPs across the mean field also other
mechanisms, such as large-scale drifts \citep{Dalla2013,Marsh2013} and
propagation along meandering field lines \citep{GiaJokMaz2000,
  Laitinen2013} have received recent attention.

Each of the suggested mechanisms would result in different type of
interplanetary transport, which may be seen in the velocity dispersion
pattern. However, without taking the diffusion in the interplanetary
medium into account in any way, it may be difficult to discern between
the different mechanisms. In this work, we derived an estimate for the
delay time of SEPs due to the scattering, using a simple model for the
onset evolution, and SEP transport simulations. We showed that the
estimate improved the VDA fitting results in the case where particles
propagate along the Parker spiral field. We tested the model by
analysing a solar event that could be assumed to be magnetically
well connected from the Sun to Earth and found that the correction brought
both the injection time and path length to be consistent with the
expected solar and interplanetary conditions. A comparison of our
model results against multi-spacecraft observations, as well as events
modelled with cross-field diffusion, drifts and meandering fieldlines,
may bring more light into the mechanism behind the efficient spreading
of SEPs in the inner heliosphere.

\section{Conclusions}\label{sec:conclusions}

In this work, we have studied the validity of the velocity dispersion
method in estimating the injection times of SEPs by using
  simulations of energetic protons. We find that the typical method
of determining the onset time as time when the intensity of the SEPs
exceeds the pre-event background in statistically significant amount
can lead to significant errors when estimating the solar injection
time and path length with the VDA method. It is important to note that
a reasonable path length does not imply a good estimate for the
injection time.

We studied the use of the intensity increase timescale, $\tau_I$ to
improve the VDA estimate. The gradual increase of the SEP intensities
in time is caused by the scattering of the SEPs in the interplanetary
space, and we find a relation between $\tau_I$, the diffusion
timescale $\tau_D$ and the delay of the SEPs with respect to the
scatter-free propagation time of the particles from Sun to the
Earth. Using this relation, we showed that the injection time estimate
given by the VDA can be improved in case of magnetically
well-connected SEP events.

We conclude that the injection times and path lengths obtained by
using the VDA method should be used with care. The interplanetary
scattering of particles does delay the arrival of the first-observed
SEPs. This delay is likely to cause errors in injection analysis based
on the observed SEP onset times particularly if the pre-event particle
intensities are high compared to the event maximum intensity.
While our analysis was performed using energetic protons, our
  results are valid for any particle species that are affected by
  interplanetary scattering, including heavy ions and electrons.

  \acknowledgements{We acknowledge support from the UK Science and
    Technology Facilities Council (STFC) (grants ST/J001341/1 and
    ST/M00760X/1). The CME catalog is generated and maintained at the
    CDAW Data Center by NASA and The Catholic University of America in
    cooperation with the Naval Research Laboratory. SOHO is a project
    of international cooperation between ESA and NASA.}

\end{document}